\documentclass{article}%
\usepackage{amsmath}
\usepackage{amsfonts}
\usepackage{amssymb}
\usepackage{graphicx}%
\providecommand{\U}[1]{\protect\rule{.1in}{.1in}}

\begin{document}

\title{On Pastewka \& Robbins' criterion for macroscopic adhesion of rough surfaces }
\author{M.Ciavarella\\Politecnico di BARI. \\V.le Gentile 182, 70125 Bari-Italy. \\Email mciava@poliba.it}
\maketitle

\begin{abstract}
Pastewka \& Robbins (PNAS, 111(9), 3298-3303, 2014) recently have proposed a
criterion to distinguish when two surfaces will stick together or not, and
suggested it shows a large conflict with asperity theories. It is found that
their criterion corresponds very closely to the Fuller and Tabor asperity
model one when bandwidth $\alpha$ is small, but otherwise involves a rms
ampliture of roughness reduced by a factor $\sqrt{\alpha}$. Therefore, it
implies  the stickiness of any rough surface is the same as that of the
surface where practically all wavelength components of roughness are removed
except the very fine ones, which is perhaps counterintuitive. The results are
therefore very interesting, if confirmed. Possible sources of approximations
are indicated, and a significant error is found in plotting the pull-off data
which may improve the fit with Fuller and\ Tabor. However, \ still they show
finite pull-off values in cases where both their own criterion and an asperity
based one seem to suggest non stickiness, and the results are in these
respects inconclusive.

\end{abstract}


\bigskip

\section{Introduction}

Pastewka \& Robbins (2014, PR in the following) recently suggested a criterion
to distinguish when two surfaces will stick together (i.e. when the area-load
curve bends into the tensile quadrant), which seems based only on fine scale
quantities like rms slopes or curvatures, and argued that it conflicts with
the classical criterion obtained by Fuller \& Tabor (1975, FT in the
following) using an asperity model, where instead emphasis is on rms
amplitude, both for stickiness and for the value of pull-off. With beautiful
atomistics simulations, PR introduce self-affine fractal roughness from a
lower wavelength $\lambda_{s}$ of order nanometers $\lambda_{s}/a_{0}%
=4,8,32,64$, to an upper wavelength $\lambda_{L}$ in the micrometer to
millimeter range, e.g. $\lambda_{L}=2048a_{0}$, where $a_{0}$ is atomic spacing.

Their main initial experiment is described as varying the level of adhesion
and adjusting the external load $N$ $=N_{rep}-N_{att}$ as to keep constant the
repulsive contact area. They find that:-

1) There is always a linear relation between the external load $N$ and the
area in intimate repulsive contact, $A_{rep}$. A result that was shown to be
robust in asperity models and was not questioned until people started to be
interested in very large bandwidths roughness. \ Defining Nayak bandwidth
parameter $\alpha=\frac{m_{0}m_{4}}{m_{2}^{2}}\sim\zeta^{2H}$, where $m_{n}$
are the moments of order n in the random process, $\zeta=\lambda_{L}%
/\lambda_{s}$ is magnification factor, and $H$ is Hurst exponent, PR systems
have for the low fractal dimensions ($H=0.8$) a Nayak $\alpha$ of the order of
1600, which is very large, and at these large bandwidths asperities coalesce
and form bigger objects which are difficult to be defined by random process
theory (Greenwood, 2007). This leads in asperity models to an area-slope which
is linear only asymptotically at large separations, and decreasing with
$\alpha^{1/4}$ otherwise (Carbone \&\ Bottiglione, 2008); but let us not
distract the reader with this point which, in the asperity adhesive models,
may tend to \textit{decrease }stickiness, whereas we shall see that PR
criterion introduces a bandwidth dependence which strongly \textit{increases}
stickiness, and with $\alpha^{1/2}$.

2) They find the attractive forces have little effect on the detailed
morphology of the repulsive contact area, suggesting the corresponding
repulsive force and mean pressure are also nearly unchanged. This suggests
they are close to the Derjaguin-Muller-Toporov (DMT) limit for which the
repulsive pressure is unaffected by adhesive forces and hence the deformation
is principally due to the repulsive forces, which in the DMT theory is given
by Hertz theory.

3) They notice that in the "attractive" regions, the pressure is simply the
theoretical strength of the material, $\sigma_{th}=w/\Delta r$, where $w$ is
surface energy, and $\Delta r$ is a range of attraction. This suggests a sort
of Dugdale-Maugis model for adhesion which requires only the knowledge of the
size of the region of attractive forces, $A_{att}$. $A_{att}$ is found to be a
fixed proportion of the repulsive one $A_{rep}$, by considering the first
order expansion of the separation distance between two contacting bodies under
repulsive forces only, which scales as distance$^{3/2}$, and equating the peak
separation to the characteristic distance $\Delta r$. Notice in particular
both $A_{att}$,$A_{rep}$ are written as a function of a perimeter $P$,
respectively\footnote{With this definition of $d_{rep}$, which is the mean
over contiguous segments in horizontal or vertical slices through $A_{rep}$ ,
for a set of $n$ circular objects, we get $A_{rep}=nd_{rep}^{2}$, instead of
$n\frac{\pi}{4}d_{rep}^{2}$ , which means that the representative diameter is
a little smaller than the real one, $d_{rep}=\frac{\sqrt{\pi}}{2}d$.}
$A_{rep}=Pd_{rep}/\pi$ and $A_{att}=Pd_{att}$, where $d_{rep}$ and $d_{att}$
are the characteristic contact diameter and the additional size of attractive
region, respectively, suggesting the contact area is a "fractal", which
requires special attention. However, at least in the limit of low bandwidths,
the simpler model of circular contact areas of diameter $d_{rep}$, and
circular annuli $d_{att}$ around the repulsive contact areas, should be
sufficient. An asperity model would also show this if it predicts the
repulsive and adhesive loads to be proportional each to the number of
asperities in contact, $n$. This will be shown to be indeed the case. In other
words, the perimeter $P$ can be given by $n\pi d_{rep}$ and the entire set of
results continues to hold for the asperity model too. For the circular area
case, in particular, the PR calculation leads to a circular attractive annulus
of size $d_{att}$ $=\frac{1}{d_{rep}^{1/3}}\left(  \frac{3}{2}R\Delta
r\right)  ^{2/3}$and an attractive load per asperity
\begin{equation}
N_{att,asp}=\pi d_{rep}d_{att}\frac{w}{\Delta r}=3^{2/3}\pi wR\left(
\frac{\delta}{\Delta r}\right)  ^{1/3}\label{N1}%
\end{equation}
where $\delta$ is the compression of the asperity, suggesting this model
doesn't lead exactly to the DMT model for a sphere (see Maugis, 2000) as
usually it is reported that for DMT the adhesive load on the asperity is
independent on its compression and is equal to $N_{att,asp}=2\pi Rw,$ the pull
off load. However, this point doesn't change the main results of this
discussion, and we shall take the PR model for the calculation of the asperity
theory, rather than the original DMT.

4)\ A condition for stickiness is found in their eqt.10
\begin{equation}
\frac{h_{rms}^{\prime}\Delta r}{\kappa_{rep}l_{a}}\left[  \frac{h_{rms}%
^{\prime}d_{rep}}{4\Delta r}\right]  ^{2/3}<1 \label{PR_condition}%
\end{equation}

In loose terms, PR criterion says nothing new: that for macroscopic bulk
solids, adhesion at the macroscale is observed only in the case of very soft
bodies of very smooth and clean surfaces, so that the length scale
$l_{a}=w/E^{\ast}$ is sufficiently large compared to $a_{0}$, where $E^{\ast}$
is plane strain elastic modulus of the material pairs, and $a_{0}$ is atomic
spacing. More precisely, there is a limit in vacuum for perfectly clear
surfaces of crystalline solids, $l_{a}/a_{0}\approx0.05$ for a Lennard-Jones
potential whose interaction distance $\Delta r\simeq a_{0}$. However, it is
the detail that matters. Using well esthablished results $\kappa_{rep}%
\approx2$, $\Delta r\approx a_{0}$ but grouping the variables using the Nayak
bandwidth parameter, we can restate\ (\ref{PR_condition}) as
\begin{equation}
\frac{h_{rms}^{\prime}a_{0}}{2l_{a}}\left[  \frac{h_{rms}}{a_{0}\sqrt{\alpha}%
}\right]  ^{2/3}<1 \label{PR_parameter}%
\end{equation}
and therefore really the condition is on rms amplitude also for PR. Despite
this condition does not correspond immediately to the original FT parameter
(which contains a radius of asperities), we shall find that a very close
equation is obtained also with very simple asperity models, except that the
$\sqrt{\alpha}$ reduction of $h_{rms}$ is not obtained, which means that
asperity models predict a much stronger reduction of stickiness with roughness amplitude.

\section{A simple asperity model}

We can restate the basic results of the FT model in a simpler form if we
consider some simplified assumptions, without changing the results
qualitatively. We consider therefore an exponential distribution $\phi
=\frac{C}{\sigma_{s}}\exp\left(  -\frac{z_{s}}{\sigma_{s}}\right)  $
($z_{s}>0$), and use the PR model for the behavior of each of the asperities
(\ref{N1}), namely the adhesive load on each asperity of radius $R$ is
dependent on compression $\delta$ with a power-law. Repeating the standard
calculation of asperity models (see Johnson, 1985), \ and the contact area as
being purely given by the compressive actions, the number of asperities (per
unit area) in contact $n$, and the total area $A$ are unchanged with respect
to the standard Hertzian case without adhesion,
\begin{align}
n &  =D_{0}\exp\left(  -\frac{d_{0}}{\sigma_{s}}\right)  \label{n}\\
A_{rep}/A_{0} &  =\pi R\sigma_{s}n\label{area}%
\end{align}
where $D_{0}$ is total number of asperities per unit area. The total load per
unit area is instead changed as%
\begin{equation}
N/A_{0}=n\left(  E\left(  \sigma_{s}^{3}R\right)  ^{1/2}\sqrt{\pi}%
-3^{2/3}\Gamma\left(  \frac{4}{3}\right)  \pi wR\frac{\sigma_{s}^{1/3}}{\Delta
r^{1/3}}\right)  \label{load}%
\end{equation}

\bigskip PR suggest a critical importance of geometry of the contact not being
"euclidean", but being fractal. They find the contact area as a intricate
geometry having a characteristic size which they estimate from purely
geometrical considerations
\begin{equation}
d_{rep}=4h_{rms}^{\prime}/h_{rms}^{\prime\prime}\label{drep-local}%
\end{equation}
which has to be multiplied by a perimeter, where the dependence on the contact
load enters. We try to reinterpret this result in the light of asperity model
maintaining circular contact areas, and simply stating that the perimeter
varies with number of asperities in contact, and is therefore a multiple of
$d_{rep}$ itself. Dividing (\ref{area}) by (\ref{n}), we have an estimate of
the mean diameter for the asperity model
\begin{equation}
\overline{d}_{rep,am}=2\sqrt{R\sigma_{s}}%
\end{equation}
and hence seems to be dependent on non-local quantities, in contrast with
(\ref{drep-local}). However, using well known quantities in random process
theories (see Carbone \&\ Bottiglione, 2008) for the product $R\sigma_{s}%
D_{0}=\frac{1}{48}\sqrt{\frac{3}{\pi}\left(  \alpha-0.9\right)  }$ which was
in early days considered to be constant, but which instead varies with
bandwidth, and for $D_{0}=\frac{1}{6\pi\sqrt{3}}\frac{m_{4}}{m_{2}}$, we get
\begin{equation}
\overline{d}_{rep,am}=2\sqrt{\frac{6\pi\sqrt{3}}{48}\sqrt{\frac{3}{\pi}\left(
\alpha-0.9\right)  }}\frac{\sqrt{m_{2}}}{\sqrt{m_{4}}}=\left(  3.2\div
10.3\right)  h_{rms}^{\prime}/h_{rms}^{\prime\prime}\label{dmean}%
\end{equation}
changing bandwidth in the range used by PR (16 to 1600), so this evaluation
gives radius generally higher than PR finds. Exact coincidence occurs only for
$\alpha\simeq40$. This is still a correct order of magnitude result with
respect to PR calculation, and indeed PR suggest that their factor 4 is an
estimate \textit{"deviations by up to a factor of 2 from this expression for
}$d_{rep}$\textit{ are responsible for the spread in the figure 3"}, but
should we attribute the scatter to a bandwidth dependence as the asperity
model predicts? It is extremely important as this assumption changes quite
radically the result on the stickiness parameter. Indeed, they also suggest
\textit{"For a given system, changes in }$d_{rep}$\textit{ with }$A_{rep}%
$\textit{ are less than 25\% over 2--3 decades in }$A_{rep}$\textit{". }Since
for a given system implies a given bandwidth, PR also find indirectly that the
most part of the variation is due to bandwidth, and the factor 2 they find
seems surprisingly in agreement with our estimate for (\ref{dmean}) which is
indeed a factor 2 larger for large bandwidths. If we were to modify their
criterion (\ref{PR_parameter}) with this $\alpha^{1/4}$ increase of $d_{rep}%
$\textit{ } with bandwidth in (\ref{dmean}), we would already restrict their
result as
\begin{equation}
\frac{h_{rms}^{\prime}a_{0}}{2l_{a}}\left[  \frac{h_{rms}}{a_{0}\alpha^{1/4}%
}\right]  ^{2/3}<1
\end{equation}

Returning to the load equation (\ref{load}), it results from a difference, and
hence it becomes zero when the contact becomes "sticky". To compare with more
advanced random process theory based asperity models (see e.g. Carbone
\&\ Bottiglione, 2008), the term $\sqrt{\frac{R}{\sigma_{s}}}$ transforms into
a slope parameter (we are confusing of course here $\sigma_{s}$ to a rms
amplitude), and therefore there is a \textit{sharp} distinction between sticky
and nonsticky behaviour\footnote{In the\ FT model, the transition is not so
sharp, but at low enough, the pull-off is so small and the region of negative
loads is obtained at so high separations, that we can consider the cases
non-sticky.} when
\begin{equation}
\chi=0.33\frac{a_{0}}{l_{a}}h_{rms}^{\prime}\left(  \frac{h_{rms}}{a_{0}%
}\right)  ^{2/3}<1 \label{Chi_parameter}%
\end{equation}
which is remarkably close both qualitatively and quantitatively to PR
parameter (\ref{PR_parameter}) at low bandwidths: exact coincidence would be
obtained for a special bandwidth parameter, which in our crude estimate is of
the order of $\alpha=3.5$.

PR criterion also suggests that, if we consider a full self-affine spectrum of
roughness, since the rms slopes and curvatures are defined only by the fine
scale features, if these fine scales satisfy the criterion, it does not matter
if we have this fine roughness structure as part of a much wider bandwidth of
roughness, or in itself. In other words, if we start of with a fine roughness
structure so that $\alpha_{fine}=2$ (fig.1b) then we can enlarge the roughness
without limit if $h_{rms,big}=\sqrt{\alpha_{big}}\frac{h_{rms,fine}}{\sqrt{2}%
}$ as in Fig.1a.

Notice that in a sense, this "removal" of large scale roughness was done by FT
in a much crude way in the sense that they had a macroscopic form, and
microscopic roughness, although it is unclear how many scales of roughness
they had in the microscopic scale. In their comparison with experiments, they
used the reduction of pull-off with respect to the case of aligned asperities
in a case like Fig.1b, for their spheres. When they did compute the adhesion
parameter, they only considered rms amplitude of the fine scale roughness:
however, they used this reduction factor to correct the pull-off value
expected for the spheres, which scales with their radius. Hence, it would seem
that in the more complex problem with multiscale roughness, if PR criterion is
correct, we expect that pull-off cannot be dependent only on this new adhesion parameter.

\begin{center}
$%
\begin{array}
[c]{cc}%
{\includegraphics[
height=2.3146in,
width=3.6611in
]%
{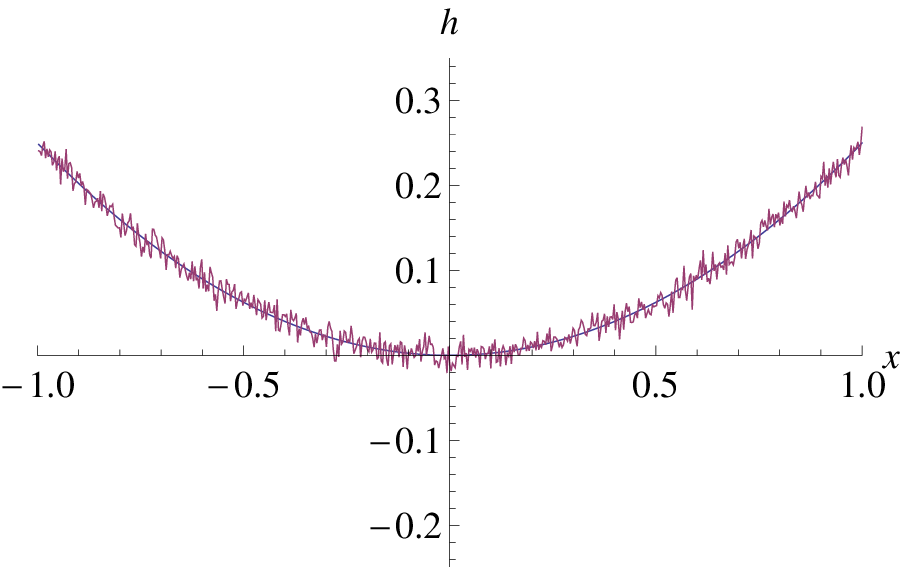}%
}
& (a)\\%
{\includegraphics[
height=2.3146in,
width=3.6611in
]%
{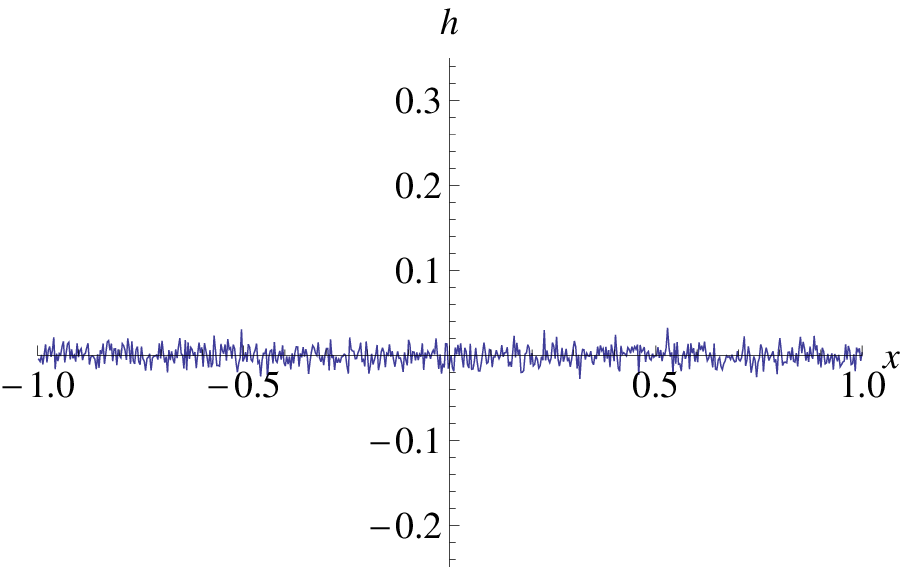}%
}
& (b)
\end{array}
$

Fig.1 An example of the "stickiness" equivalence in PR criterion (a) a local
fine scale roughness, on a larger wavelength structure of which we show only
some parts, (b) the same roughness but now in itself.
\end{center}

\section{\bigskip Pull-off}

PR have also interesting data for pull-off in their "Supplementary
Information", which they find in error with respect to the FT prediction by
several orders of magnitude and also qualitatively not in good order. First,
we should note an \textit{error} in the scale of their Fig.S3. PR were aiming
at using the scale used by FT, the ratio of pull-off load to sum of pull-off
of aligned total number of asperities, $\frac{N_{\max}}{N_{tot}2\pi Rw}%
$\footnote{PR use $\frac{3}{2}\pi Rw$ for a single asperity as in JKR theory,
instead of DMT value which may be more appropriate, but this is irrelevant.}%
but they assumed $R\sigma_{s}D_{0}=0.05$ which was correct in the old days for
low bandwidths\ (it is $\frac{1}{48}\sqrt{\frac{3}{\pi}\left(  \alpha
-0.9\right)  }$, which is 0.05 only for $\alpha=7$ ) whereas they bandwidth
spans the range $\alpha=16-1600$. So, if we keep their points as they are, we
should have many curves for FT, spanning a band. For the largest bandwidths,
the FT curves would be almost 2 orders of magnitude higher. Some points may
still too "sticky" than what FT predicts, and especially "stickiness" results
for a much wider range than the original FT adhesion parameter, in agreement
with the main difference we found in the stickiness parameters. However, the
reduction on rms amplitude needed to collapse the data in the x-axis is at
most a factor 2, whereas the new PR criterion suggests a much higher
reduction, scaling with $\sqrt{\alpha}$. \ PR suggest that their data are the
"lower bound"\ of pull-off forces they can find, since these are
load-dependent. Would other pull-off forces be closer to a FT theory
"corrected" with the new adhesion parameter? It is impossible without
estimating all individual bandwidths in the data.

However, yet another contradiction appears from the data: the caption says
$h_{rms}^{\prime}=0.1$ or $0.3$ (closed and open symbols), $l_{a}/a_{0}=0.005$
(blue) or $l_{a}/a_{0}=0.05$ (red). Hence, their own criterion now reads with
$h_{rms}^{\prime}=0.1$ and the case with low adhesion $l_{a}/a_{0}=0.005$,
suppose with $\lambda_{S}=4a_{0}$, and $\alpha=1600$
\begin{equation}
\frac{h_{rms}}{a_{0}}<\left(  \frac{1}{10}\right)  ^{3/2}\sqrt{1600}=1.26!
\end{equation}
and by no means their surfaces are so small in rms amplitude to be of atomic
size. Indeed, the rms amplitude they have can be estimated in this case to be
$h_{rms}=h_{rms}^{\prime}\lambda_{s}\left(  \lambda_{L}/\lambda_{s}\right)
^{H}=0.1\ast4a_{0}\left(  1000\right)  ^{0.8}=\allowbreak100a_{0}$. Even more
absurd with $\lambda_{S}=64a_{0}$, bandwidth $74$, still with same parameters,
PR criterion reads $\frac{h_{rms}}{a_{0}}<\left(  \frac{1}{10}\right)
^{3/2}\sqrt{74}=0.27$. If we take now $l_{a}/a_{0}=0.05$, these numbers will
be multiplied by 10, which doesn't solve the problem: most point in the plot
should be non-sticky \textit{both for their criterion and an asperity based
one}. These pull-off values correspond, in the correct scale, to the pull-off
of a relevant number of asperities out of the total number, and do not seem to
be plausible with the parameters of roughness they have.

\section{Discussion}

We have pointed out that the "parameter-free" theory of PR may contain several
important approximations which affect the stickiness criterion. In particular,
their assumption of a constant factor 4 in $d_{rep}$ (\ref{drep-local}) seems
problematic even in fig.3 PR show, and conflicts by the same factor as we have
estimated in an asperity model. The fact that asperity models at low
bandwidths correctly describe the geometry of the problem is well accepted
today (Greenwood, 2007), so the source of conflicts seems to be the dependence
on $\alpha$ for large bandwidth. 

Another difference with the asperity model is hidden in assuming mean values
for both diameter of repulsive contact area and size of annulus of attraction.
In PR parameter-free theory, $d_{re\not p}$ is the mean diameter and does seem
to take into account of the distribution of contact spot sizes, and so does
$d_{att}$. In fact, the asperity model does not need to make this
approximation. If I estimate the mean size of the annulus of attraction
directly from the $\overline{d}_{rep,am}$ in (\ref{dmean}), as $\overline
{d}_{att}$ $=\frac{1}{d_{rep}^{1/3}}\left(  \frac{3}{2}R\Delta r\right)
^{2/3}$, I\ get
\begin{equation}
\frac{A_{rep}}{A_{att}}=\frac{n\pi\overline{d}_{rep,am}^{2}/4}{n\pi
\overline{d}_{rep,am}\overline{d}_{att}}=\frac{\left(  \sigma_{s}%
/a_{0}\right)  ^{2/3}}{4\left(  3/2\right)  ^{2/3}}\simeq\frac{\left(
\sigma_{s}/a_{0}\right)  ^{2/3}}{5.24}%
\end{equation}
whereas if I estimate $\frac{A_{rep}}{A_{att}}$ from the full integration
process which takes into account of the distribution of contact spots sizes
(\ref{load}) as%
\begin{equation}
\frac{A_{rep}}{A_{att}}=\frac{N_{att}/A_{rep}}{w/a_{0}}=\frac{\left(
\sigma_{s}/a_{0}\right)  ^{2/3}}{3^{2/3}\Gamma\left(  \frac{4}{3}\right)
}\simeq\frac{\left(  \sigma_{s}/a_{0}\right)  ^{2/3}}{1.85}%
\end{equation}
which suggests a 3 times less area of attraction. As $\alpha^{1/4}$ varies
from 16$^{1/4}=\allowbreak2.0$ to 1600$^{1/4}=\allowbreak6.\,\allowbreak3$,
this factor 3 is not irrelevant. It may well be that these sublte differences
in the factor are better captured by the PR model instead of the asperity
model, but the result seems quite counterintuitive. 

\section{Conclusion}

We have reexamined the results of PR recent "parameter-free" theory. The
parameter-free theory in fact does contain some parameters, and in particular,
the estimate of the diameter of the repulsive contact areas, which deserves
further attention. Despite asperity theories are known to be possibly in error
at large bandwidth parameters, many results PR find numerically do not seem in
conflict, except of course the criterion for stickiness, which corresponds
only in the limit of low bandwidths.  The new criterion contains a curious
implication, that one can take some fine scale roughness, and build on it
increasingly larger wavelengths of roughness without affecting the stickiness.
Since a finite stickiness implies also a finite pull-off, this seems to be an
interesting results, which requires further proof. Unfortunately, the data
they present for pull-off do not seem consistent, and do not permit conclusive discussion.

\section{References}

Carbone, G., \& Bottiglione, F. (2008). Asperity contact theories: Do they
predict linearity between contact area and load?. Journal of the Mechanics and
Physics of Solids, 56(8), 2555-2572.

Fuller, K. N. G., \& Tabor, D. (1975). The effect of surface roughness on the
adhesion of elastic solids. Proc Roy Soc London A: 345, No. 1642, 327-342

Greenwood, J. A. (2007). A note on Nayak's third paper. Wear, 262(1), 225-227.

Johnson, K. L., K. Kendall, and A. D. Roberts. 1971. Surface energy and the
contact of elastic solids. Proc Royal Soc London A: 324. 1558.

Johnson, K.L., (1985). Contact Mechanics. Cambridge University Press.

Maugis, D. (2000). Contact, adhesion and rupture of elastic solids (Vol. 130).
Springer, New York.

Pastewka, L., \& Robbins, M. O. (2014). Contact between rough surfaces and a
criterion for macroscopic adhesion. Proceedings of the National Academy of
Sciences, 111(9), 3298-3303.

\end{document}